\documentclass[aps,twocolumn,twoside,nofootinbib,prd,preprintnumbers]{revtex4}
\usepackage{graphicx}
 
\newcommand{\lsim}{\lesssim}
\newcommand{\gsim}{\gtrsim}
\newcommand{\ra}{\rightarrow}

\begin{document}

\bibliographystyle{revtex}

\title{Four Generations, the Electroweak Phase Transition, and Supersymmetry}

\author{Ricky Fok}
\affiliation{\mbox{Department of Physics and Institute of Theoretical Science,
University of Oregon, Eugene, OR 97403}}
 
\author{Graham D. Kribs}
\affiliation{\mbox{Department of Physics and Institute of Theoretical Science,
University of Oregon, Eugene, OR 97403}}

\begin{abstract}

We calculate the strength of the electroweak phase transition 
in a supersymmetric model with four chiral generations.
The additional chiral fermions (and scalar partners) lower the 
critical temperature and thus strengthen the first-order phase transition.  
The scalar partners stabilize the potential, leading to an effective
theory that is bounded from below.  We identify the ensemble of
parameters where $\phi_c/T_c \gsim 1$ simultaneous with obtaining
a large enough Higgs mass.  Our calculations focus on a subset 
of the full four generational supersymmetric parameter space:
We take the pseudoscalar heavy, $\tan\beta=1$, and neglect all
subleading contributions to the effective potential.
We find that the region of parameter space with a strong first-order 
phase transition requires $m_{\tilde{q}'}/m_{q'} \lsim 1.1$ while the 
constraint on the lightest Higgs mass requires $m_{\tilde{q}'}/m_{q'} \gsim 1$ 
with $m_{q'} \gsim 300$ GeV\@.  We are led to an intriguing prediction
of quarks and squarks just beyond the current Tevatron direct search limits
that are poised to be discovered quickly at the LHC\@.

\end{abstract}

\maketitle

\section{Introduction}
\label{intro-sec}

The origin of the matter asymmetry is a deep mystery that
remains unsolved.  Conditions that can lead to a dynamical asymmetry 
between baryons and anti-baryons were articulated years ago
by Sakharov \cite{Sakharov:1967dj}:  baryon number violation, 
C and CP violation, and a departure from thermal equilibrium.
All three conditions are satisfied by the Standard Model
as it passes through the electroweak phase transition.
But, the CP violation is too small \cite{Jarlskog:1985ht},
and the phase transition is not strongly first-order
(e.g., \cite{Anderson:1991zb,Dine:1991ck,Arnold:1992rz,Cohen:1993nk,Quiros:1999jp}),
given the direct search bounds on the Higgs boson from LEP\@.
New physics with large CP violation is trivial to introduce 
into the model; weak scale supersymmetry is an obvious example 
(care needs only to be taken to ensure that induced electric 
dipole moments are within the experimental bounds).  
Even with a new source of CP violation, if the phase 
transition is not strong enough first-order, any generated 
baryon asymmetry will be washed out \cite{Kuzmin:1985mm}.
New physics that enhances 
the first-order phase transition, however, is generally much more 
tricky to achieve.  

In the early 1990s it was realized that the electroweak phase
transition could be enhanced by modifying
the cubic coupling of the finite-temperature effective potential 
\cite{Anderson:1991zb}.
Nontrivial modifications of the cubic coupling could arise from additional 
scalars with order one couplings to the Higgs.  In the
minimal supersymmetric standard model (MSSM), the scalar 
superpartners to the top quarks (stops) can play precisely this role
\cite{Giudice:1992hh,Espinosa:1993yi,Carena:1996wj}.  It has long 
been advocated that the region of MSSM parameter space with a light stop 
(and a light Higgs) can yield a strong enough phase transition.  
Unfortunately, the combination of direct searches for the 
lightest Higgs boson and direct searches for stops have virtually ruled out 
this possibility.  The remaining parameter space \cite{Morrisseytalk}
requires a large hierarchy between the left-handed and right-handed
stops to ensure the Higgs satisfies the LEP bound.

Methods to strengthen the first-order phase transition beyond the SM
and MSSM are now widely discussed
\cite{Pietroni:1992in,Kang:2004pp,Carena:2004ha,Menon:2004wv,Ham:2004nv,Funakubo:2005pu,Huber:2006wf,Shu:2006mm,Profumo:2007wc}.
Several of these ideas add a singlet field, such as in the NMSSM or 
nMSSM\@.  Another related idea is to simply cutoff the SM at a low scale,
adding the effects of higher dimensional operators \cite{Grojean:2004xa}
(which can be UV completed by integrating out a singlet).

Yet another interesting possibility, and the one we will focus on 
in this paper, is to add more particles with modestly strong couplings 
to the Higgs.  
This was proposed in \cite{Carena:2004ha}; the additional particles 
have quantum numbers such that they mix with the MSSM charginos 
and neutralinos.
Heavy particles that receive their mass entirely or dominantly
from electroweak symmetry breaking can have a substantial
impact on the electroweak phase transition.
In this paper we consider a modification to the MSSM similar in 
spirit to \cite{Carena:2004ha} to enhance the phase transition.  
Namely, we add a fourth generation 
of particles (and sparticles) to the MSSM\@. 
Larger couplings to the
Higgs are automatic simply due to the direct search bounds from 
LEP and Tevatron on fourth generation fermions.

That the electroweak phase transition could be enhanced in a 
four generation supersymmetric model was considered before 
in \cite{Ham:2004xh}.  
They performed an interesting numerical study that also found that 
the electroweak phase transition can be enhanced when 
$m_{\tilde{q}'}/m_{q'}$ is not much larger than 1.  
However, the present limits on fourth generation quark masses 
rule out their parameter space, and moreover, they allowed 
$\tan\beta$ to far exceed 1, implying 
that the $b'$ Yukawa coupling was nonperturbative.  In our analysis, 
we first systematically analyze the origin of the contributions that 
allow the phase transition to become first-order.  This allows us to make 
a clear distinction between supersymmetric and non-supersymmetric 
theories with heavy chiral fermions.  We then identify the viable region 
of parameter space where the Yukawa couplings are under control 
($\tan\beta \simeq 1$) and all other bounds are satisfied.

A fourth generation has historically been thought be strongly
disfavored by the absence of flavor mixing, the $Z\ra \nu\bar{\nu}$
constraint, and electroweak precision data.  All of these objections
can be straightforwardly overcome, as was recently emphasized 
in \cite{Kribs:2007nz}.  Below we summarize these results in the form 
of the parameter space that is allowed.
It is interesting that the constraints from electroweak
precision data can be overcome without or with an electroweak
scale Majorana mass for the fourth generation right-handed neutrinos.
If a Majorana mass does indeed exist, lepton number is violated 
at the electroweak scale, and thus scenarios of baryogenesis that rely 
on an earlier generation of $B-L$ number (such as leptogenesis) 
do not work here \cite{Fukugita:1990gb,footnoteibe}.
Electroweak baryogenesis 
is one of the few mechanisms not sensitive to this source
of lepton number violation, and thus becomes even more interesting
to study.

\section{Setup}

We consider a low energy supersymmetric theory with a fourth
chiral generation of matter, the ``4MSSM''
(for an early discussion, see \cite{Gunion:1994zm}).
A fourth chiral generation of matter does affect electroweak
precision observables.  One of the main results of \cite{Kribs:2007nz}
was to show that it can be made completely consistent with 
electroweak precision data so long as there are modest mass 
splittings between the isospin partners in the quark and lepton 
doublets.  This splitting causes a modest reduction in the positive
contribution to S, simultaneous 
with a positive contribution to T,
allowing ensembles of parameters that are within the 68\% CL
ellipse of the LEP Electroweak Working Group.  For example,
the mass spectrum $m_{\nu'} = 100$ GeV, $m_{\ell'} = 155$ GeV,
$m_{t'} = 310$ GeV, $m_{b'} = 260$ GeV, and $m_h = 115$ GeV 
is perfectly acceptable.  There is strong sensitivity to the
mass \emph{differences} with only mild sensitivity to the 
overall scale of the particles.  We will present results for
both the electroweak preferred ratio $m_{t'}/m_{b'} \sim 1.2$ as well as
$m_{t'} = m_{b'}$ for comparison.

With supersymmetry, there are additional contributions to 
electroweak precision observables from superpartners 
(e.g., \cite{Pierce:1996zz}).
We have not included these contributions to optimize the
parameter set to match electroweak data, simply because
many more parameters enter the fit that can be freely adjusted
without affecting our results for the electroweak phase
transition.  We therefore take fourth generation Yukawa couplings
consistent with \cite{Kribs:2007nz} and take the scalar
partner masses to be degenerate, eliminating this potential
additional contribution to isospin violation.

We neglect all sub-leading contributions to the 
zero-temperature and finite temperature effective potential.
Sub-leading here refers to couplings smaller than about 1.
We retain the contributions from gauge bosons.
But we neglect light fermions (u,d,c,s,b,e,$\mu$,$\tau$), 
Higgs bosons (the quartic is small), and all superpartners 
other than $\tilde{t'}_{1,2}, \tilde{b'}_{1,2}, \tilde{t}_{1,2}$.
We also neglect contributions from fourth generation leptons
because the number of degrees of freedom per particle is only
$1/3$ that of quarks and the bounds on the mass from the 
non-observation in experiment are much weaker than for quarks.

\section{Supersymmetry with $\tan\beta = 1$}

In the limit $\tan\beta \ra 1$, several aspects of supersymmetry
drastically simplify.  From the definition of 
$\tan\beta \equiv \langle H_u \rangle / \langle H_d \rangle$,
we see the vevs are equal, $v_u = v_d = v/\sqrt{2}$, where
$v = 246$ GeV, and the Yukawa couplings
\begin{eqnarray}
y_f = \frac{2 m_f}{v} \; ,
\end{eqnarray}
are the same for the up-type and down-type fermions.
The fourth generation quarks $t'$ and $b'$ have large Yukawa
couplings,
\begin{eqnarray}
y_{t',b'} = 2.1 \left( \frac{m_{t',b'}}{260 \; {\rm GeV}} \right) \; ,
\end{eqnarray}
where $260$ GeV is shown for convenience in comparison
to the (approximate) present direct search bounds from the Tevatron 
\cite{:2008nf,d0-higgs-fourth}.
Note that these Yukawa couplings are a factor of $\sqrt{2}$ 
larger than in a non-supersymmetric model, since the $t'$ 
and $b'$ acquire their mass only through couplings to the 
up-type and down-type Higgs, respectively.  

\begin{figure}[t]
\centering
\includegraphics[width=0.45\textwidth]{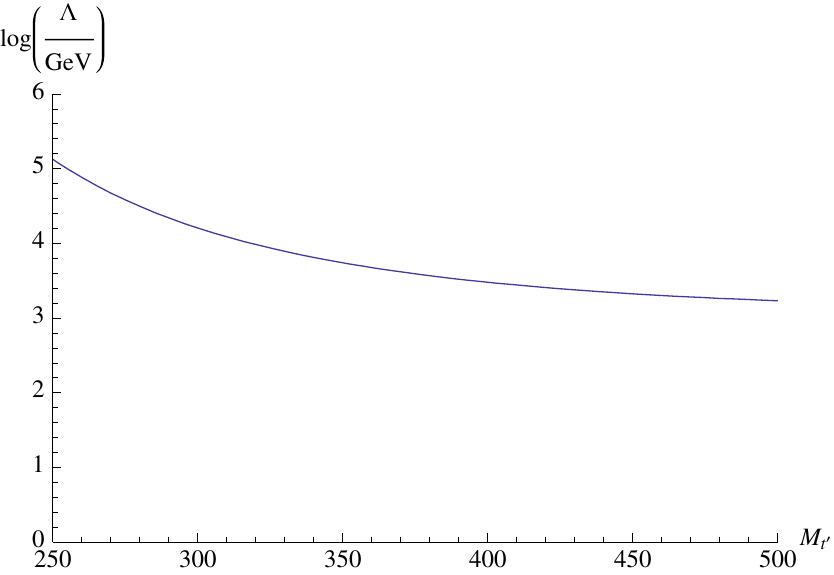}
\caption{An estimate of the cutoff scale of the 4MSSM
as a function of the fourth generation fermion mass 
by running the one-loop RGEs of the quark Yukawas up to where 
they encounter a Landau pole ($y_{f'} \sim 4\pi$).}
\label{landaupoles-fig}
\end{figure}

If $\tan\beta \not= 1$, either the up-type
or down-type Higgs vev is reduced, and thus to hold the masses
of the fermions fixed, either $y_{t'}$ or $y_{b'}$ must increase.
Setting $\tan\beta = 1$ allows the largest possible 
physical fourth generation fermion masses with the smallest 
Yukawa couplings.  Since $y_f$ cannot be arbitrarily large 
for perturbative calculations at the weak scale to be valid, 
the parameter choice
$\tan\beta = 1$ really just maximizes the cutoff scale 
of the model.  Even with this adjustment, the cutoff scale
is low.  This can be estimated by running the one-loop 
renormalization group equations for the Yukawa couplings 
up to $\sim 4\pi$.  We show the scale of the Landau pole 
as function of fermion mass in Fig.~\ref{landaupoles-fig}.
Note that requiring $y_f^2/(4 \pi) \lsim 1$, 
implies $y_f \lsim 3.5$, corresponding to $m_f \lsim 450$ GeV;
we will not consider fermion masses that much exceed this value.

In the limit $\tan\beta \ra 1$ the Higgs sector also drastically
 simplifies.  The tree-level potential in the MSSM with 
$\tan\beta = 1$ is \cite{Martin:1997ns}
\begin{eqnarray}
\lefteqn{V = \left( m_{H_u}^2 + \mu^2 \right) |H_u^0|^2 
+ \left( m_{H_d}^2 + \mu^2 \right) |H_d^0|^2} \nonumber \\ 
& & \hspace*{-4mm} - (b H_u^0 H_d^0 + {\rm c.c.})
+ \frac{1}{8} \left( g^2 + g'^2 \right) 
\left( |H_u^0|^2 - |H_d^0|^2 \right)^2 .
\label{higgspot-eq}
\end{eqnarray}
Expanding the neutral components as
\begin{equation}
\left( \begin{array}{c} H_u^0 \\ H_d^0 \end{array} \right) =
\frac{1}{\sqrt{2}}
\left( \begin{array}{c} v \\ v \end{array} \right) +
\frac{1}{\sqrt{2}}
\left( \begin{array}{cc}  \cos\alpha & \sin\alpha \\ 
                         -\sin\alpha & \cos\alpha \end{array} \right)
\left( \begin{array}{c} h \\ H \end{array} \right) 
\end{equation}
where the $\alpha$ rotation matrix diagonalizes the Higgs fluctuations 
($h$, $H$) into mass eigenstates.
In the limit that the second Higgs doublet ``decouples''
($m_{A^0,H,H^\pm} \gg m_h$), the mixing angle $\alpha \ra -\beta$, 
and thus the lightest Higgs is simply $h = (H_u^0 + H_d^0 - \sqrt{2} v)$.  
In this limit the tree-level Higgs potential vanishes, since $h$
corresponds to the excitation of a $D$-flat direction.  

Since electroweak precision data prefers $m_{t'}/m_{b'} \simeq 1.2$, 
this could be arranged either by adjusting just these two Yukawa
couplings $y_{t'}/y_{b'} = 1.2$ or instead adjusting
$\tan\beta = 1.2$.  These two scenarios are nearly equivalent
for our purposes, and so we choose to set $\tan\beta = 1$.
The alternative, $\tan\beta = 1.2$, would give a tree-level contribution
to the Higgs potential.  The contribution to the (mass)$^2$ is however
just $0.03 M_Z^2$.
As we will see, the one-loop radiative corrections from quarks
and squarks will be far larger than this, 
so it is safe to completely neglect tree-level contributions
even if $\tan\beta$ were allowed to vary slightly from $1$.

In addition to taking $\tan\beta = 1$, we also choose supersymmetric 
parameters such that the mass eigenstates of $\tilde{t'}_{1,2}$ 
and $\tilde{b'}_{1,2}$ correspond to the gauge eigenstates 
$\tilde{t'}_{L,R}$ and $\tilde{b'}_{L,R}$.  This is done
purely to simplify our calculation.  It is a rather conservative
approximation, since it is well known that increasing
the off-diagonal contribution to the squark mass matrix
leads to an enhancement in the one-loop contribution to
the Higgs mass (e.g., see \cite{Carena:2000dp}).
We expect that the parameter space with
a strong first-order phase transition will enlarge as 
this restriction is relaxed.
Note that since the off-diagonal left-right contribution to the 
up-type and down-type squark mass matrix is equal to 
$m_f (A_f - \mu)$ (where again, $\tan\beta = 1$), this 
simplification corresponds to the specific parameter 
choice $A_f = \mu$.

Finally, as we discussed above, the Higgs potential simplifies
in the limit $m_{A^0}, m_{H^\pm}, m_{H^0} \gg m_h$.
This is a common assumption in the electroweak phase transition
literature:  The operational advantage is that the low energy 
theory is effectively a one-Higgs-doublet model that is drastically
simpler to analyze at finite temperature.

\section{One-loop effective potential}

In the 4MSSM with $\tan\beta = 1$, loop corrections entirely determine
the Higgs potential.  We are interested in the loop corrections to
just the scalar fluctuation $\phi = (h + v)/\sqrt{2}$.
At one-loop the effective potential for the Higgs is determined 
from the Coleman-Weinberg potential
\begin{eqnarray}
V_1 &=& \sum_i \frac{n_i}{64 \pi^2} M_i(\phi)^4 
\left( \ln \frac{M_i(\phi)^2}{\mu^2} - c_i \right)
\label{cw-pot-eq}
\end{eqnarray}
where $M_i(\phi)$ are the field-dependent masses and $\mu$ is the
renormalization scale ($\overline{\rm MS}$ scheme).
We generally use $M_i$ to refer to $\phi$-dependent 
(and temperature-dependent) masses and $m_i$ to refer to $M_i(v)$
at zero temperature.
The $c_i$'s are constants corresponding to $5/6$ for gauge bosons
and $3/2$ for fermions and scalars.  The degeneracies per particle
are $n_q = -12$ (for each $q = t,t',b'$), 
$n_{\tilde{q}_L} = n_{\tilde{q}_R} = 6$, $n_{W_T} = 4$, 
$n_{Z_T} = 2$, $n_{W_L} = 2$, $n_{Z_L} = 1$.

Expanding the effective potential as given above, evidently the minimum
is not necessarily located at the proper electroweak breaking scale
$v = 246$ GeV\@.  This is easily remedied by imposing a
renormalization condition on the mass parameter 
such that the minimum is enforced to be at $v$.  
This amounts to adding the $v$-dependent contribution to 
the effective potential,
\begin{eqnarray}
\lefteqn{\Delta V = - \frac{d V_1(\phi=v)}{d v^2} \phi^2 \; =} \nonumber \\ 
& & \hspace*{-4mm} 
- \sum_i \frac{n_i}{32 \pi^2} M_i^2(v) \frac{d M_i^2(v)}{d v^2} 
\left( \ln \frac{M_i^2(v)}{\mu^2} \! + \! \frac{1}{2} \! - \! c_i \! \right) 
\phi^2 .
\label{mass-pot-eq}
\end{eqnarray}

The masses used in the effective potential are $\overline{\rm MS}$ masses
that differ from the physical (pole) masses through finite and
log-dependent corrections.  The running fermion masses are given at
one-loop by
\begin{eqnarray}
\left. m_f \right|_{\rm pole} &=& m_f(\mu) \left[ 
1 + \frac{\alpha_s}{\pi} 
\left( \frac{4}{3} + \ln \frac{\mu^2}{m_f^2} \right) \right] \; .
\end{eqnarray}
Since the fourth generation fermions overwhelmingly dominate the 
contributions to the (zero- and finite-temperature) effective potential, 
we take $\mu = \sqrt{m_{t'} m_{b'}}$, 
i.e., the scale of the largest electroweak breaking masses in the problem. 
This tends to minimize the higher order corrections to the potential,
though are calculations are not particularly sensitive to the
precise choice of renormalization scale.

The running scalar masses also differ from their physical pole masses
through one-loop corrections depending on not only the gluon but also
gluino diagrams \cite{Pierce:1996zz}.  This correction is
generally numerically smaller than the correction to the fermion mass,
typically less than a few \%.  
Moreover, since the correction is gluino mass-dependent, 
relating the pole mass to the running mass requires specifying
an otherwise unfixed parameter in our model.
We choose instead to simply take 
$m_{\tilde{f},{\rm pole}} = m_{\tilde{f}}(m_{\tilde{f}})$,
thus neglecting the difference between the pole and running mass
for the squarks.

\section{Finite temperature one-loop effective potential}

\begin{figure*}[t]
\centerline{
\includegraphics[width=0.5\textwidth]{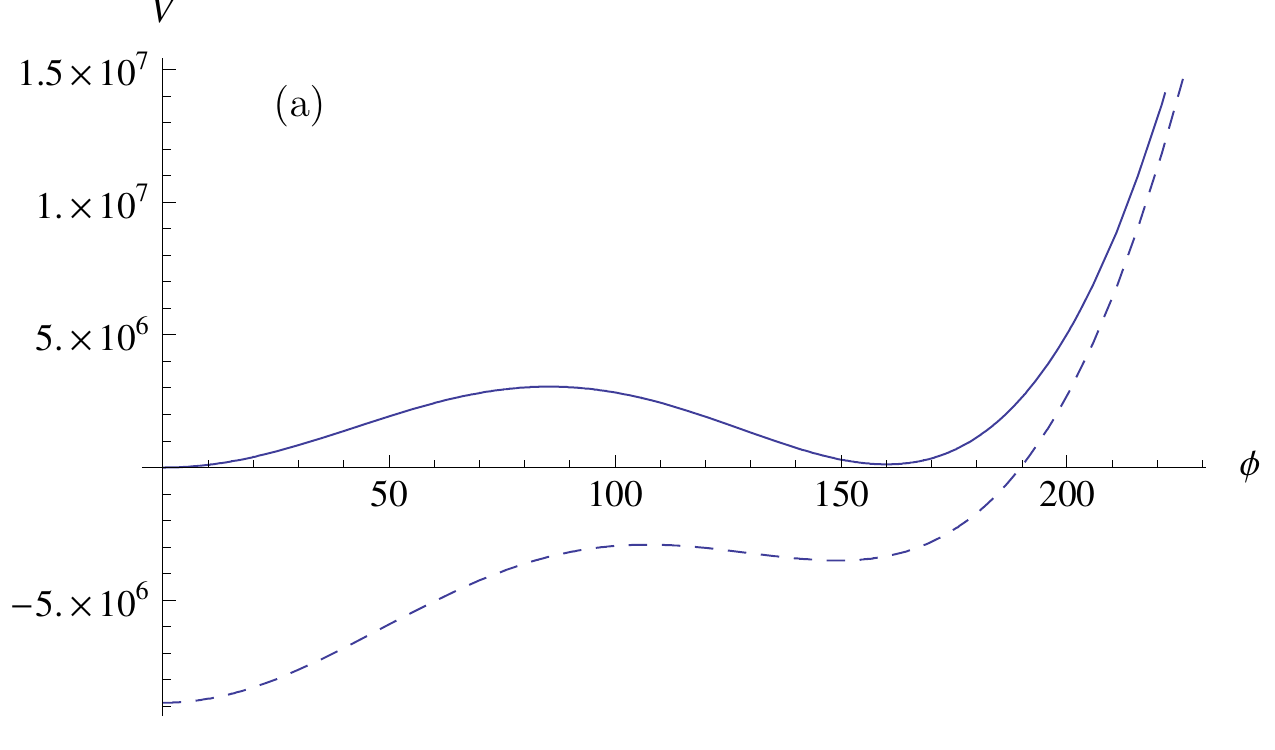}
\hspace*{0mm}
\includegraphics[width=0.5\textwidth]{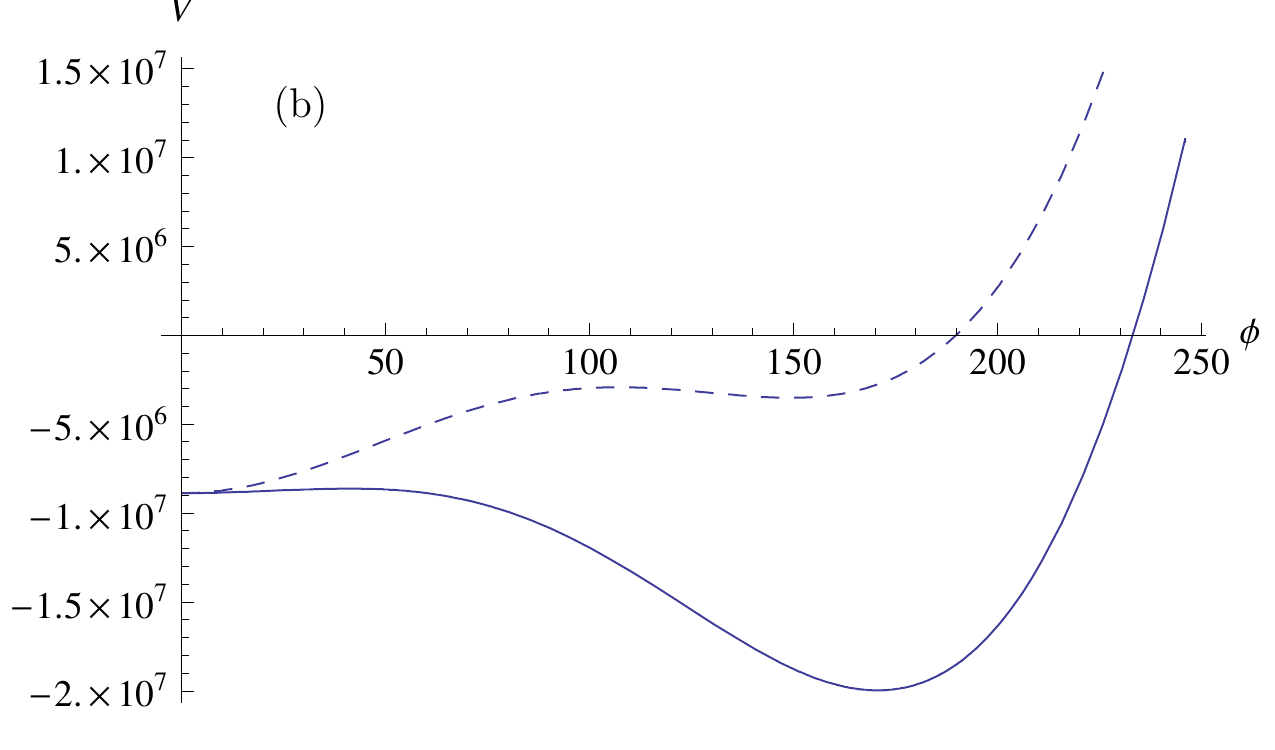}}
\caption{An illustration of the effect of adding one additional
heavy fermion that obtains a mass of 400 GeV entirely from 
electroweak symmetry breaking.  The figure (a) on the left shows 
the effective potential at the critical temperature $T_c$ 
(solid line) in the 4MSSM with $m_{t'} = m_{b'} = 400$ GeV
and $m_{\tilde{q}'}/m_{q'} = 1.05$.  Adding an additional 
heavy fermion (with $n = -1$) to the effective potential 
is shown (dashed-line), except that \emph{only} the finite-temperature 
contribution, $V_{T1}$, is included in (a).  Figure (b) on the right shows 
the the effect of including just the finite-temperature contribution
(dashed-line), identical to Figure (a), and then the effect
of including both the finite-temperature as well as the 
zero-temperature contribution $V_1$ (solid-line).  The net effect 
shown in Figure (b) solid-line is that the global minimum at 
$\phi \simeq \phi_c$ decreases and thus $T_c$ increases.}
\label{heavyfermion-fig}
\end{figure*}

The finite-temperature contributions to the effective potential are 
\cite{Anderson:1991zb,Dine:1991ck,Arnold:1992rz,Cohen:1993nk,Quiros:1999jp}
\begin{eqnarray}
V_T &=& V_{T1} + V_{\rm ring} 
\end{eqnarray}
where
\begin{eqnarray}
V_{T1} &=& \sum_i \frac{n_i}{2 \pi^2} J_i\left(\frac{M_i^2}{T^2}\right) T^4 \\
V_{\rm ring} &=& - \frac{T}{12 \pi} \sum_{k=W_L,Z_L} n_k 
\left( \bar{M}_k^3 - M_k^3 \right) \label{ring-eq}
\end{eqnarray}
and 
\begin{eqnarray}
\bar{M}_k^2 &=& M_k^2 + \Pi_k \; .
\end{eqnarray}
The field-dependent fermion and scalar masses are
\begin{eqnarray}
M_f(\phi)^2 &=& 2 y_f^2 \phi^2 \\
M_{\tilde{f}}(\phi)^2 &=& m_{\rm soft}^2 + M_f(\phi)^2 \; .
\end{eqnarray}
Explicit expressions for the thermal masses of the SM gauge
bosons can be found in, e.g., Ref.~\cite{Carena:1997ki}.
The finite-temperature contributions depend on the thermal functions
\begin{equation}
J_{B,F}(y^2) = \int_0^\infty d x \, x^2 \ln 
\left[ 1 \mp \exp \left( - \sqrt{x^2 + y^2} \right) \right] .
\end{equation}
Often a high temperature approximation is employed to estimate
these integrals.  In our case, due to the large Yukawa couplings,
this approximation is generally not appropriate.  Consequently, 
all computations given below evaluate the thermal functions 
$J_{B,F}$ numerically.  
The ring contribution (\ref{ring-eq}) is only relevant for the
longitudinal components of the $W$ and $Z$.  No contributions 
from scalars are included here since the squarks receive a 
contribution from soft supersymmetry breaking, and thus they 
remain heavy in the $\phi \ra 0$ limit.

We have self-consistently included contributions in the finite-temperature
effective potential the same as those included in the zero-temperature
effective potential.  Namely, we include contributions from $t',b',t$, 
their superpartners $\tilde{t}'_{L,R},\tilde{b}'_{L,R},\tilde{t}_{L,R}$,
as well as the transverse and longitudinal components of $W,Z$.  

In the standard model, the electroweak 
phase transition becomes second order when the
Higgs mass approaches 70 GeV \cite{Laine:1998jb}.
Qualitatively, this is because the transverse modes of $W$ and $Z$, 
which would drive a first-order phase transition in the standard model
with a lighter Higgs mass, 
develop a thermal mass from non-perturbative effects. If
the transverse thermal masses are large, they effectively remove the cubic term
from the finite-temperature potential when the effective potential 
is reset to zero at $\phi = 0$.  In our model, the first-order 
phase transition is mostly driven by squarks. In fact, when the $W$ and $Z$
are neglected in our model, $\phi_c$ and $T_c$ does not change significantly
and the phase transition remains first-order. Therefore, we do not expect 
non-perturbative effects encountered in the standard model at larger
Higgs masses to significantly affect our calculations of the strength 
of the phase transition in the 4MSSM\@.

\section{Effects of new heavy particles}

The effects of heavy particles (that receive their mass dominantly 
from electroweak symmetry breaking) on the electroweak phase transition
can be broadly characterized as follows.
Consider the effective potential at $T_c$, where there are two 
degenerate minima $V_{eff}(0,T_c) = V_{eff}(\phi_c,T_c)$ located
at $\phi = 0$ and $\phi = \phi_c$.
Now add to this a new particle that satisfies $M(\phi_c)/T_c \gg 1$. 
The phase transition strength is modified in two ways from 
the contributions of the new particle.  
One is through corrections to the finite-temperature
contribution; the other is through the zero-temperature 
Coleman-Weinberg potential.

\subsection{Finite-temperature effects}

The contributions from bosons and fermions with masses larger than 
the critical temperature, $m \gg T_c$, can be characterized by 
how they contribute at large field values $\phi \gg T$ and
small field values $\phi \ll T$.  At large field values, we can take
a \emph{low temperature} approximation to the finite-temperature effective
potential.  In this limit, the contribution from fermions or bosons 
becomes
\begin{eqnarray}
\left. V_{T1} \right|_{T \ll m} &=& -|n| \left( 
\frac{M(\phi)}{2 \pi T} \right)^{3/2} T^4
\exp \left[ -\frac{M(\phi)}{T} \right]
\label{low-t-eq}
\end{eqnarray}
where $n$ counts the number of degrees of freedom per boson
or fermion with field-dependent mass $M(\phi)$.  
Clearly, when $M(\phi) \gg T$, which is equivalent to 
$\phi \gg T$ (with order one or larger Yukawa couplings),
the contribution to the effective potential from fermions or bosons
is exponentially suppressed.  

At small field values, we can take a high-temperature approximation
to the thermal contribution to the effective potential.
The leading order contribution is the field-independent constant
\begin{eqnarray}
\left. V_{T1} \right|_{T \gg m} &=& -|n| c_{B,F} \frac{\pi^2}{90} T^4
\label{high-t-eq}
\end{eqnarray}
where $c_{B,F} = (1,7/8)$ for a boson or fermion contribution.

The combination of (\ref{low-t-eq}) and (\ref{high-t-eq}) imply
that the introduction of a heavy fermion or boson causes a 
substantial negative shift in the potential at $\phi = 0$
while causing a negligible shift in the potential at $\phi = \phi_c$.
As an illustration, we show in Fig.~\ref{heavyfermion-fig}(a)
the effect of adding one additional heavy fermionic degree of freedom
($n_f = -1$, for illustration)
that obtains a mass of 400 GeV entirely from electroweak symmetry breaking.  
Readjusting the minimum $V_{\rm eff}(\phi = 0) = 0$ shifts the
potential up for all field values, thereby removing the second minimum
at $\phi = \phi_c$, and thus restoring electroweak symmetry.
We must \emph{lower} the temperature further in order to have the
second minimum reappear in the effective potential with the new
heavy fermion or boson.

\subsection{Zero-temperature Effects}

The second effect of heavy bosons and fermions is that they
also modify the zero-temperature effective potential.  
Here, however, the effect of fermions and bosons is different.
There are two contributions whose origin is ultimately
the Coleman-Weinberg potential.  One contribution is to
the quartic coupling (\ref{cw-pot-eq}), while the second
contribution is the quadratic term (\ref{mass-pot-eq}).
For smaller field values, i.e., $\phi \lsim \mu$, the
dominant contribution is from the quadratic term.
Since we choose $\mu \simeq m_{q'}$, the log term drops out
of (\ref{mass-pot-eq}),
giving an overall \emph{negative} (\emph{positive}) 
contribution to the effective potential from fermions (bosons).

\begin{figure}[t]
\centering
\includegraphics[width=0.48\textwidth]{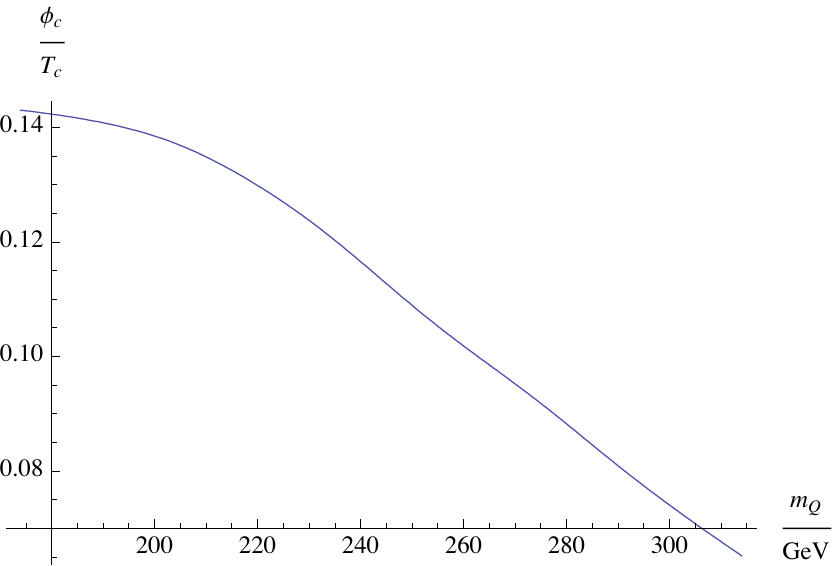}
\caption{This plot shows that the electroweak phase transition 
order parameter decreases if an additional heavy chiral quark is 
added to the theory without its corresponding superpartner.}
\label{newfermion-fig}
\end{figure}

The negative contribution from fermions at modest field values
actually overpowers the effect from the finite-temperature contributions
discussed above.  This is illustrated in Fig.~\ref{heavyfermion-fig}(b).
The net result is that introducing a new heavy 
chiral fermion causes a \emph{decrease} in $\phi_c/T_c$ as the 
mass of the fermion is increased.  The resulting decrease in the 
strength of the electroweak phase transition with one additional 
chiral fermion is shown in Fig.~\ref{newfermion-fig}.

Adding bosonic contributions cancels the contribution from fermions
in the Coleman-Weinberg potential.  This cancellation is one-loop exact 
in the limit $m_{\tilde{f}'} = m_{f'}$, i.e., no SUSY breaking contribution
to the scalar mass.  This makes it clear that we need 
both heavy fermions \emph{and} scalars with equal numbers of 
degrees of freedom and similar masses to utilize the mechanism 
of Ref.~\cite{Carena:2004ha} to lower $\phi_c/T_c$.

\begin{figure}[t]
\centering
\includegraphics[width=0.48\textwidth]{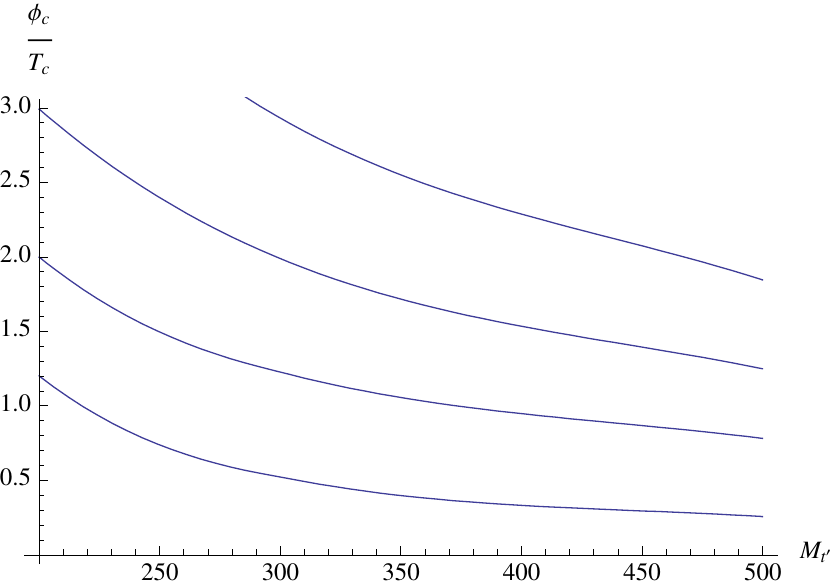}
\caption{Contour plot with the ratio $m_{\tilde{q}'}/m_{q'}$ fixed, 
from top to bottom, as $1.05, 1.1, 1.15, 1.2$.  The masses of the 
fourth generation quarks are taken to be equal.}
\label{MSSMorderpar}
\end{figure}

Ref.~\cite{Carena:2004ha} estimated that of order ten or more 
degrees of freedom is needed to enhance the phase transition 
sufficiently to achieve $\phi_c/T_c \gsim 1$.  A fourth generation
quarks corresponds to adding 24 degrees of freedom.  
(We could have equivalently added degrees of freedom
in other ways, such as several pairs of vector-like supersymmetric 
lepton doublets that only get mass through the Higgs mechanism.
This is another interesting possibility that we will not
explore here \cite{footnoteroy}.) 
We have calculated the strength of the phase transition for a range
of quark and squark masses.  The results are shown in 
Fig.~\ref{MSSMorderpar}.

\section{Lightest Higgs mass in the 4MSSM}

\begin{figure}
\centering
\includegraphics[width=0.48\textwidth]{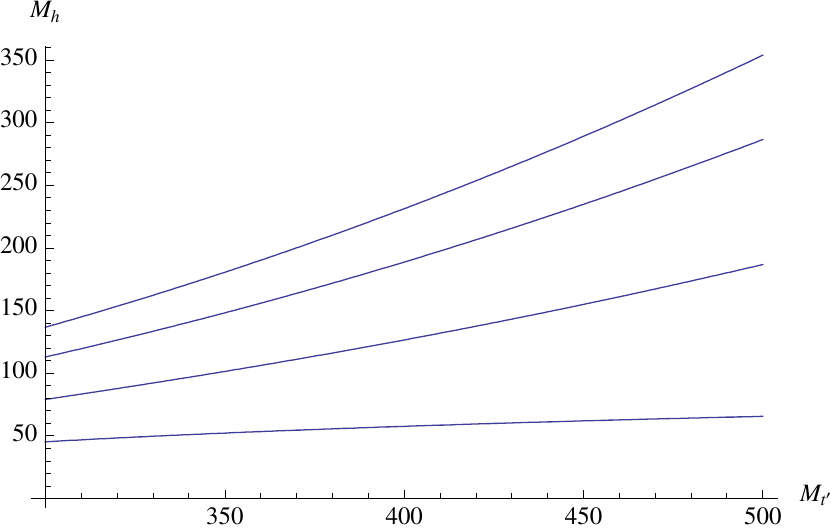}
\caption{Contour plot of the lightest Higgs mass as a function
of the fourth generation quark mass.  Each contour corresponds
to a fixed ratio $m_{\tilde{q}'}/m_{q'}$, from top to bottom, 
1.2, 1.1, 1.0, 0.95 with $m_{t'} = m_{b'}$ and 
$m_{\tilde{t}} = m_{\tilde{q}'}$.}
\label{MSSMHiggsmass-fig}
\end{figure}

Given the parameter choice $\tan\beta = 1$, the tree-level Higgs 
potential vanishes, and thus the lightest Higgs mass 
also vanishes at leading order.  It is well known that loop corrections 
from splitting the masses of the top quark from the stops 
in the Coleman-Weinberg potential 
provides large corrections to the tree-level value.
In the 4MSSM, we can split not only the top and stops, but also split 
the fourth generation quarks from squarks.  Since the one-loop contribution 
to the Higgs quartic coupling is proportional to $y_{f}^4$, even a small 
splitting between $f$ and $\tilde{f}$ has a very important effect
on the Higgs mass.
A one-loop estimate of the lightest Higgs mass in the 4MSSM
can be obtained by taking $d^2(V_0+V_1)/d\phi^2$ at $\phi = v$.
This gives our rough estimate for the Higgs mass
\begin{eqnarray}
m_h^2 &=& \sum_{f=t,t',b'} \frac{3}{2 \pi^2} 
\frac{m_{f}^4}{v^2} \ln \frac{m_{\tilde{f}}^2}{m_f^2} \; .
\label{higgsmass-eq}
\end{eqnarray}
where again $v = 246$ GeV\@.  In Fig.~\ref{MSSMHiggsmass-fig}
we show Higgs mass plotted against different $m_{t'} = m_{b'}$ masses,
where all squark masses were taken to be degenerate
$m_{\tilde{t}'} = m_{\tilde{b}'} = m_{\tilde{t}}$.
Each contour has the fourth generation squark-to-quark mass ratio,
$m_{\tilde{f}'}/m_{f'}$, fixed to the values shown.  
Clearly, when the splitting between 
the fourth generation squark and quark ($\overline{\rm MS}$) 
masses vanishes, there is an 
insufficient one-loop contribution from top/stop loops to raise the 
Higgs mass much above about 60 GeV\@.  Nevertheless, for even a small
splitting between fourth generation squarks and quarks, 
one can easily obtain a one-loop contribution to the Higgs
mass that far exceeds the LEP bound so long as 
$m_f \gsim 300$ GeV\@.  

\section{Results}

\begin{figure*}[t]
\centerline{
\includegraphics[width=0.5\textwidth]{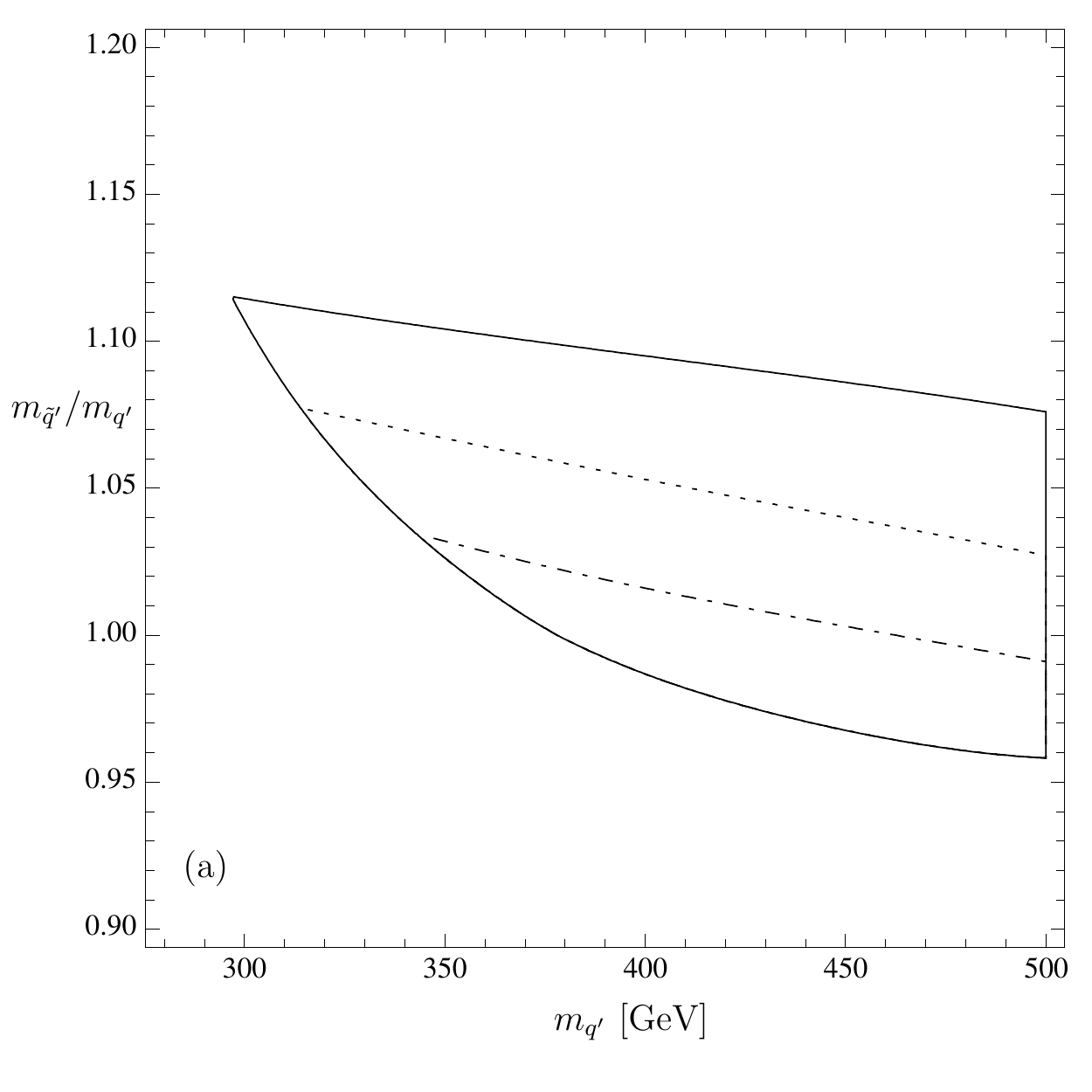}
\hspace*{0mm}
\includegraphics[width=0.5\textwidth]{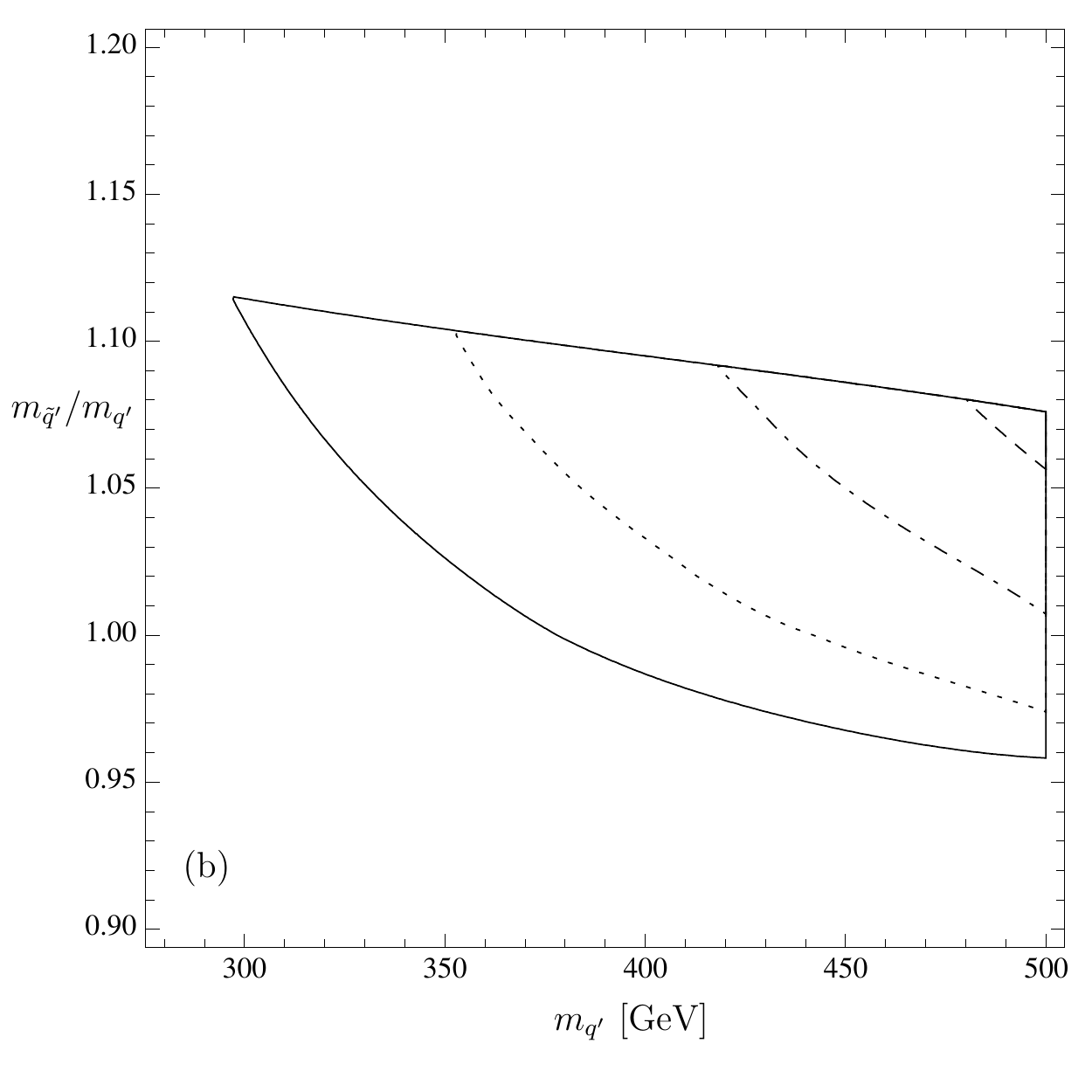}}
\vspace*{-1em}
\caption{The region in the quark/squark mass plane where the 
electroweak phase transition is strongly first-order is shown.  
The regions shown in the left- and right-hand side figures 
(a) and (b) are identical:  the upper boundary (the solid line) 
is determined by $\phi_c/T_c = 1$ while the lower boundary is 
determined by $m_h = 115$ GeV\@.  The dotted and dashed contours 
on the left-hand side figure (a) corresponding to $\phi_c/T_c = 1.5, 2.0$ 
respectively.  The dotted, dot-dashed and dotted contours on the
right-hand side figure (b) correspond to the Higgs masses 
$m_h = 150, 200, 250$ GeV respectively.}
\label{regionplot}
\end{figure*}

\begin{figure*}[t]
\centerline{
\includegraphics[width=0.5\textwidth]{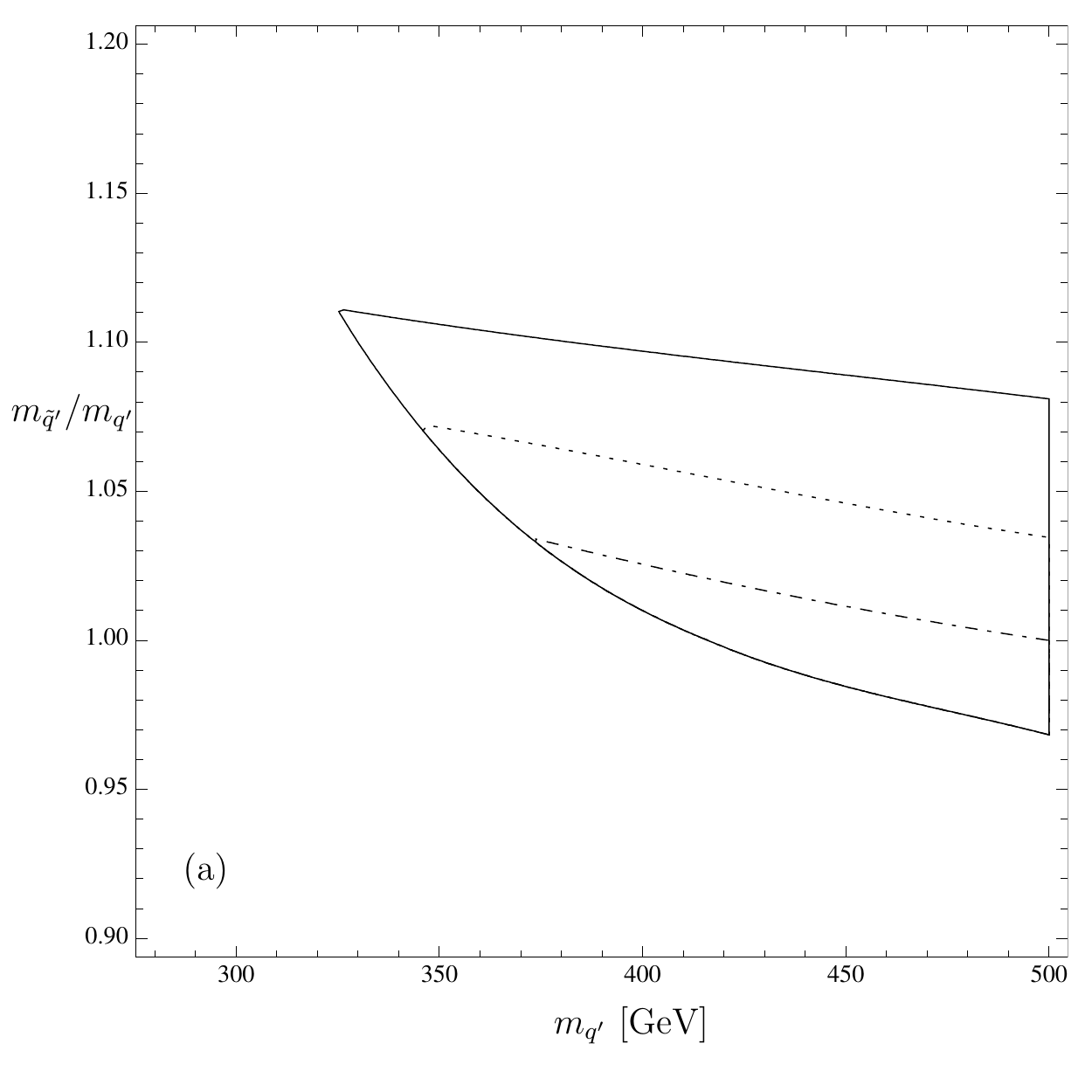}
\hspace*{0mm}
\includegraphics[width=0.5\textwidth]{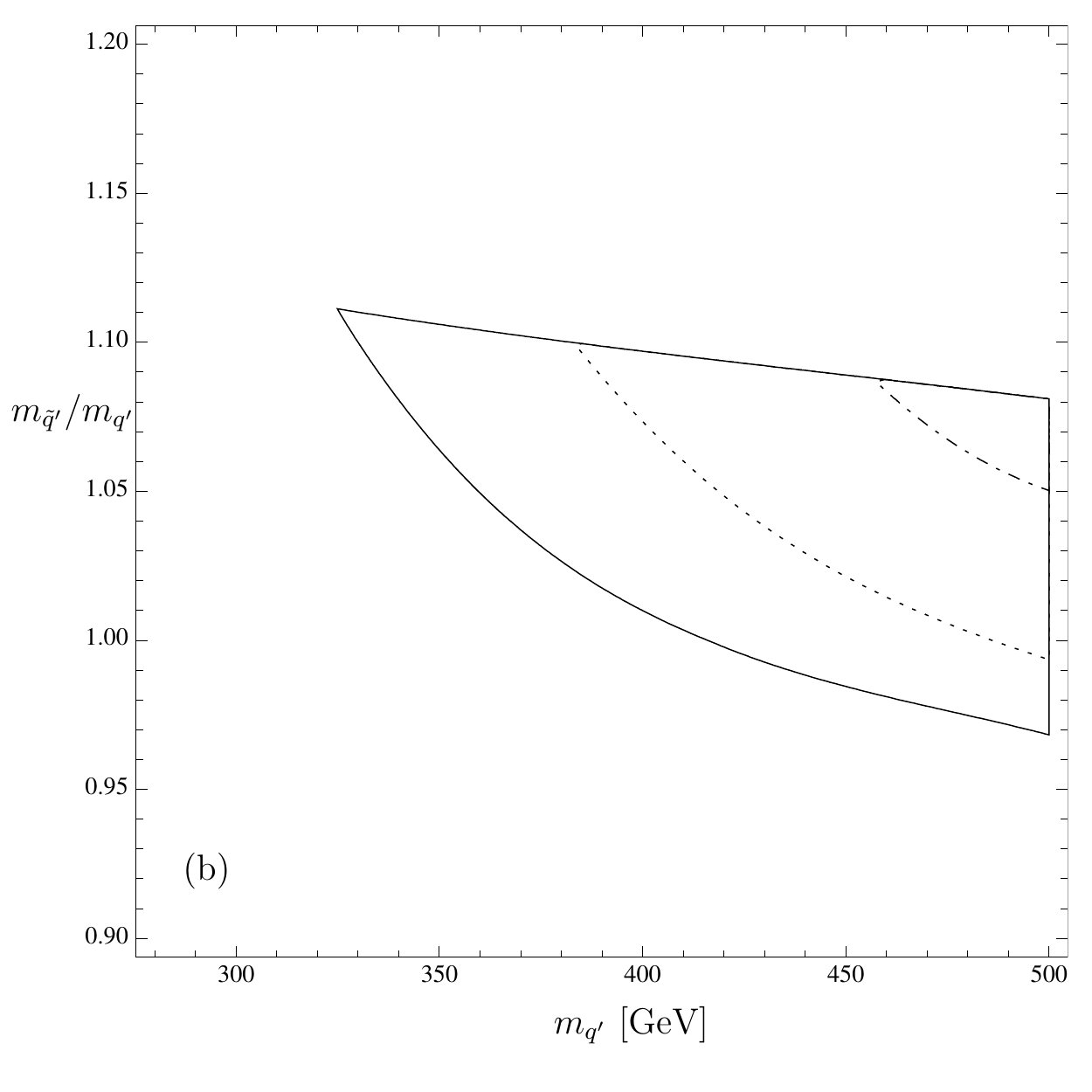}}
\vspace*{-1em}
\caption{Same as Fig.~\ref{regionplot} but we take 
$m_{t'}/m_{b'} = 1.2$, as favored by electroweak precision data. 
The basic shape and size of the region is the same, illustrating
that our results are not particularly sensitive to the heavy
fourth generation quark mass ratios.}
\label{regionplot2}
\end{figure*}

Combining our calculation of the phase transition with our
calculation of the Higgs mass, we can find the allowed region
in parameter space where the first-order phase transition
is strong $\phi_c/T_c \gsim 1$ while the Higgs mass satisfies
the LEP bound $m_h > 115$ GeV\@.  We have computed this for
the mass ratio $m_{t'}/m_{b'} = 1$ in Fig.~\ref{regionplot}
and $m_{t'}/m_{b'} = 1.2$ in Fig.~\ref{regionplot2}.  
In Fig.~\ref{regionplot}(a) we show contours of 
increasing $\phi_c/T_c$, illustrating that it is straightforward
to obtain the strength of the phase transition to 
significantly exceed $\phi_c/T_c = 1$.
In Fig.~\ref{regionplot}(b) we show contours of increasing
Higgs mass, illustrating that it is also straightforward to
to obtain a lightest Higgs mass that significantly exceeds 
$m_h = 115$ GeV\@.

Note that our plots are with respect to the pole masses
of quarks and squarks (as well as the ratio $m_{\tilde{q}'}/m_{q'}$).
The quantities that enter the effective potential are 
$\overline{\rm MS}$-renormalized masses, which differ 
(as we discussed above) for fermion masses.
Since the fermion pole mass is larger than its 
$\overline{\rm MS}$-renormalized counterpart by about 5\%, 
the ratio of pole masses can be as small as $0.95$ while the 
ratio of $\overline{\rm MS}$ masses is still larger than one.
This is why the fourth generation contributions to the Higgs mass 
(\ref{higgsmass-eq}) remains positive even when the pole mass ratio 
$m_{\tilde{q}'}/m_{q'}$ is smaller than one.  

These results suggest that even though only one-loop approximations
for the effective potential and the Higgs mass calculation
were employed, we are not near any critical boundary, and so a 
more refined calculation is expected to only modestly adjust the 
parameter regions we have shown.
For instance, there are several effects that can increase
the Higgs mass to values higher than our calculation.  One is 
moving away from the parameter choice $A_f = \mu$, where the
off-diagonal contribution from squarks provides a modest 
increase.  Another is separating the third generation squark
masses from the fourth generation squark masses.  The latter
effect is also modest:  in Fig.~\ref{MSSMHiggsmass-fig}, 
the contour $0.95$ corresponds
to just the contribution from the third generation, and one
can see that there is a small increase as the squark mass
(equal to the fourth generation quark mass) increases.
Similar statements also hold for the finite-temperature effective potential, 
since again our results show that there are model parameters where 
the 4MSSM model has an electroweak phase transition with $\phi_c/T_c$ 
that is well above the critical first-order boundary $\simeq 1$.

\section{Conclusions}

We have calculated the strength of the electroweak phase transition
in a supersymmetric model with four chiral generations.  We find
there is an intriguing region of parameter space, with fourth
generation quarks heavier than about 300 GeV and the squark to quark
mass ratio $1 \lsim m_{\tilde{q}'}/m_{q'} \lsim 1.1$,
where $\phi_c/T_c > 1$.  Within this region of parameter space
we showed the Higgs can be easily heavier than the LEP bound
of 115 GeV\@.

This suggests that a viable model of electroweak baryogenesis
could indeed be a low energy supersymmetric model with a fourth generation
of chiral fermions.
What we have shown is the the strength of the first-order 
phase transition can be large enough to prevent the washout
of a baryon asymmetry.  This model also has several new sources
of CP violation, ubiquitous in low energy supersymmetry, 
that could be used to satisfy Sakharov's CP violation criteria.
Examples of sufficient CP violation that have been employed
in other supersymmetric electroweak baryogenesis scenarios
\cite{Giudice:1992hh,Espinosa:1993yi,Carena:1996wj}
include the phase of the Higgsino mass parameter $\mu$ 
as well as the gaugino mass parameters $M_{1,2}$.  

It is \emph{coincidental} that the region of parameter space where
the first-order transition is strong enough combined with
obtaining a large enough Higgs mass (taking $\tan\beta = 1$)
happens to be just beyond the current Tevatron direct 
search bounds \cite{:2008nf,d0-higgs-fourth}.  If we are lucky, 
the Tevatron could begin to see evidence for new physics in the form of
both an extra chiral generation as well as superpartners
in the very near future.  The LHC, however, can easily 
cover this parameter space.  Indeed, the mechanism to enhance
the first-order phase transition described here is expected to
be found or ruled out with only modest amount of data from 
the LHC\@.  \\

\underline{Note added:}  
As this paper was being completed, Ref.~\cite{Hou:2008xd} appeared,
speculating that the electroweak phase transition could be
enhanced with a fourth generation \emph{without} supersymmetry.  
Unfortunately, this does not work, as we show in Fig.~\ref{newfermion-fig} 
where just adding fermions actually \emph{decreases} $\phi_c/T_c$ 
because of the effects of the fermions on the zero-temperature 
effective potential.

\begin{acknowledgments}

We thank M.~Ibe, D.~Reeb, T.~Roy, G.~Servant, T.~Tait, 
and C.~Wagner for discussions.
GDK thanks CERN for hospitality where part of this work was done.
This work was supported in part by the Department of Energy 
under contract DE-FG02-96ER40969.

\end{acknowledgments}

\end{document}